\documentclass[aip,jap,reprint,a4paper,superscriptaddress]{revtex4-1}
\usepackage{graphicx}
\usepackage{natbib}
\usepackage{amsmath}
\usepackage{amssymb}
\usepackage{color}
\usepackage{esint}
\usepackage{subfigure}
\usepackage[colorlinks=true,citecolor=blue,linkcolor=blue,hyperindex]{hyperref}
\definecolor{green}{rgb}{0,0.5,0}

\begin{document}

\title{Effect of perpendicular uniaxial anisotropy on the annihilation fields of magnetic vortices}

\author{E.R.P. Novais}
\affiliation{Centro Brasileiro de Pesquisas F\'{\i}sicas, 22290-180,  Rio de Janeiro, RJ, Brazil }

\author{S. Allende}
\affiliation{Departamento de Ciencias F\'isicas, Universidad Andr\'{e}s Bello, Avenida Rep\'{u}blica 220, 837-0134, Santiago, Chile}
\affiliation{Departamento de F\'{\i}sica, Universidad de Santiago de Chile and CEDENNA, Avda. Ecuador 3493, Santiago, Chile}

\author{D. Altbir}
\affiliation{Departamento de F\'{\i}sica, Universidad de Santiago de Chile and CEDENNA, Avda. Ecuador 3493, Santiago, Chile}

\author{P. Landeros}
\affiliation{Departamento de F\'{\i}sica, Universidad T\'ecnica Federico Santa Mar\'{\i}a, Avenida Espa\~{n}a 1680, Valpara\'{\i}so, Chile}

\author{F. Garcia}
\affiliation{Laborat\'orio Nacional de Luz S\'{\i}ncrotron, 13083-970, Campinas, SP, Brazil}
\altaffiliation[Present address: ]{Centro Brasileiro de Pesquisas F\'{\i}sicas, 22290-180,  Rio de Janeiro, RJ, Brazil}

\author{A.P. Guimar\~aes}
%\affiliation{Centro Brasileiro de Pesquisas F\'{\i}sicas, 22290-180,  Rio de Janeiro, RJ, Brazil}

\date{\today }

\begin{abstract}
The magnetic vortex structure, that is present in several nanoscopic systems, is stable and can be manipulated through the application of a magnetic field or a spin polarized current.  The size and shape of the core are strongly affected by the anisotropy, however,  its role on the core behavior has not yet been clarified. In the present work we investigate the influence of a perpendicular anisotropy on the annihilation and shape of magnetic vortex cores in permalloy disks. We have used both micromagnetic simulations with the OOMMF code, and an analytical model that assumes that the shape of the core does not change during the hysteresis cycle, known as the rigid core model, to calculate the annihilation fields. In both cases we found that the annihilation fields decrease with increasing perpendicular anisotropy for almost all the structures  investigated. The simulations show that for increasing anisotropy or dot thickness, or both, the vortex core profile changes its shape, becoming elongated. For every dot thickness, this change does not depend on the dot radius, but on the relative distance of the core from the center of the dot.
\end{abstract}

\maketitle

\section{Introduction}
\label{Sec:Introduction}
Among the nano- and mesoscopic magnetic structures that have attracted the attention of researchers in recent years stand out those that exhibit a vortex, since this state presents both interesting physical properties and a high potential for applications.\cite{Guimaraes:2009,Chien:2007,Guslienko:2008b,Bohlens:2008,Ruotolo:2009,Jung:2012}
Magnetic vortex states in nanodots are characterized by in-plane magnetic moments curling around a core which has magnetization pointing out-of-plane. Two main features are defined in a vortex,  the circulation, i.e., the sense of the magnetization curling, being $-1\, (+1)$ for clockwise (counterclockwise) rotation direction, and the polarity defined by the direction of the core magnetization denoted by  $p = +1\, (-1)$ for upward (downward) direction. The core profile $m_z(r)$ (the $z$ component of the unit magnetization) of a vortex in equilibrium is cylindrically symmetric, usually approximated by a Gaussian curve surrounded by a small dip (see Fig.~\ref{Fig:profile-h10d500Center}).\cite{Guimaraes:2009,Bode:2008}

A vortex configuration is the ground state of different nanodots with regular shape such as ellipses, squares, spheres, caps and disks, with lateral dimensions ranging from one hundred nanometers to a few microns, with some tens of nanometers  thickness.\cite{Landeros:2005,Altbir:2007,Zhang:2008,metlov:2008,Soares:2008,Chung:2010,Novais:2011}  While an external in-plane magnetic field that increases continuously from zero is applied to a disk exhibiting a magnetic vortex, its core will be displaced perpendicularly to the field direction, until its center reaches the disk edge.  The field corresponding to this limiting situation, i.e., a field that expels the vortex core, is known as the annihilation field. A further field increase will expel the vortex from the disk, and the saturated state will eventually be reached. On the other hand, when starting from a fully saturated state, by decreasing the field to a certain critical value (commonly referred in the literature as the nucleation field) the vortex will again be formed.  The knowledge and  control of the magnitude of these fields is a key issue for several applications considering the manipulation of magnetic vortices, such as non-volatile magnetic memory devices, or high-resolution magnetic field sensors. \cite{Rahm:2003,Kim:2008,Garcia:2010}

The vortex-core nucleation and annihilation processes have been discussed by several authors \cite{Schneider:2000,Fernandez:2000,Guslienko1:2001,Guslienko2:2001,Mejia:2006,Mejia:2010} and, in particular, the influence of extrinsic properties on the annihilation field has been taken into account.  Wu \textit{et al.} \cite{Wu:2008} investigated the role of geometrical asymmetries, finding that the annihilation of the vortex depends strongly on the asymmetry. The effect of the shape asymmetry has also been studied by Dumas \textit{et al.} \cite{Dumas:2009}, by measuring the angular dependence of the annihilation field.  Mihajlovi\'c \textit{et al.} \cite{Mihajlovic:2010} have shown that temperature also affects the reversal mechanism and the vortex annihilation field, while experiments by Davis \textit{et al.} \cite{Davis:2010} suggest that the nucleation and annihilation fields depend on the magnetic field sweep rate.

This problem was also examined from the theoretical point of view, within the framework proposed by Guslienko {\it et al}.\cite{Guslienko1:2001}  This model approximates the core as a magnetization distribution whose profile  does not change during the reversal process.

Some important properties of the vortices, such as the core size and some dynamic features, can be tailored introducing a uniaxial perpendicular magnetic anisotropy, as has been recently shown\cite{Garcia:2010,Machado:2008}  In this case, as the perpendicular anisotropy increases, important deviations from the vortex core profile and from the canonical magnetic vortex configuration result. Beyond a critical value of the anisotropy ($K_z^{crit}$), it is no longer observed a vortex, with the formation of a skyrmion (e.g., Fert {\it et al.}\cite{Fert:2013}), a structure that was found in experiments with BFeCoSi\cite{Yu:2010} and, more relevant to the present study, was also apparent in experiments with Co/Pt disks\cite{Garcia:2010} and simulations.\cite{Garcia:2010,Novais:2011} 

Vortex core deformations, even under the action of a magnetic field, have not been so far systematically analyzed.

The aim of this paper is to get a better understanding of the vortex annihilation process in magnetic dots.  For this, we have compared the description using the rigid vortex model with results obtained by micromagnetic simulations. In order to explore the effect of the perpendicular anisotropy on the vortex core properties, and concomitantly, verify the limits of validity of the rigid vortex model, we have introduced an anisotropic term in both the theory and simulations. We also characterized the vortex core deformations that are present in some simulations.

The paper is organized as follows: after the Introduction we describe how we perform our micromagnetic simulations that lead us to study the annihilation fields extracted from the hysteresis curves of disks with various sizes (Sec.~\ref{Sec:Simulations}).   Analytical calculations are presented in Sec.~\ref{Sec:Analytical}, with the inclusion of anisotropy terms into the rigid vortex model. The results are contained in Sec.~\ref{Sec:Results}, and finally, in Sec.~\ref{Sec:Conclusions} we summarize and draw conclusions.

\section{Numerical Simulations}
\label{Sec:Simulations}
We investigated the hysteresis loops of individual magnetic nanodots  defined by their thickness $L$ varying between 10 and 30\,nm, and diameters $D$ from 100 to 1000\,nm, while the uniaxial perpendicular anisotropy $K_z$ ranges from 0 to 300 kJ/m$^3$.  This study was conducted through micromagnetic simulations \cite{Kim:2010,Novais:2011} using the OOMMF code.\cite{oommf}  We used a stiffness constant $A=13\times 10^{-12}$\,J/m and a saturation magnetization $M_s=860\times10^3$\,A/m, the standard values used for bulk permalloy, taking a cell size of $5\times 5\times 5$ nm$^3$. The maximum anisotropies used in this work that keep the vortex structure are $K_z^{max}=300,\ 225 \  {\rm and} \  165$\,kJ/m$^3$ for the thicknesses $L=10, \ 20 \ {\rm and}\ 30$\,nm, respectively.

\begin{figure}
 \includegraphics[width=1\columnwidth]{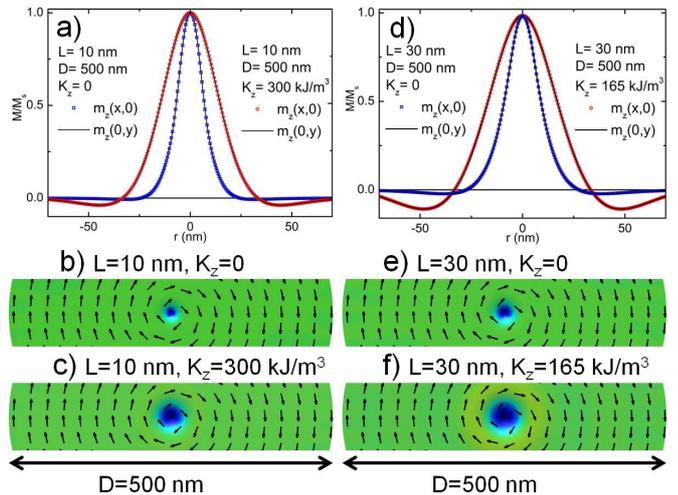}
 \caption{(Color Online) {\bf a)} profile of the vortex core corresponding to disks with  $D=500$ nm and $L=10$ nm for $K_z =0$ and $K_z=300$ kJ/m$^3$;
{\bf b)} and {\bf c)} depict the magnetization for $K_z=0$ and $K_z=300$ kJ/m$^3$ with $L=10$ nm.
{\bf d)} profile of the vortex core with  $D=500$ nm and $L=30$ nm for $K_z =0$ and $K_z =165$ kJ/m$^3$. {\bf e)} and {\bf f)} represent the magnetization for $K_z=0$ and $K_z=165$ kJ/m$^3$ for $L=30$ nm. Note that from {\bf a)} to {\bf d)} the depth of the negative part of the magnetization (the dip) increases.}
\label{Fig:profile-h10d500Center}
\end{figure}

For larger anisotropies, a skyrmion structure is observed, and perpendicular magnetization appears on the rim of the disk. For this reason, in all our calculations the anisotropy constant value was chosen such that the magnetic configuration at zero external applied field is a vortex configuration, as shown in Fig.~\ref{Fig:profile-h10d500Center}.

In our simulations we developed a systematic study of the annihilation field
 that is determined from the maximum of the derivative $dM/dB$ in the increasing magnetization branch of the hysteresis loop, which corresponds to the expulsion of the vortex core. All hysteresis curves  were obtained starting from the unperturbed configuration of the disks (with the vortex core at the center), increasing the field from $B=0$, in steps of $\Delta B=0.1$ mT, leading us to obtain the annihilation field, and finally reaching the magnetic saturation. In some simulations we  observed  a deformation of the vortex core. In order to characterize it we define
\begin{equation}
\delta =\frac{r_y-r_x}{r_x} ,
\label{eq:deformation}
\end{equation}
\noindent
where $r_x$ and $r_y$ are the sizes of the vortex core along the $x$ and $y$ axes, respectively.  As shown in Fig. \ref{Fig:profile-h10d500Center}, two orthogonal sections (x and y directions) of the profiles of the vortex core passing through the core center (maximum of $m_z$) were made. The dimensions of the core along the $x$  and $y$ directions were obtained by the full widths at half maximum of the repective profile fit, using a pseudo-Voigt function.

\section{Analytical Model}
\label{Sec:Analytical}
To obtain analytical expressions for the annihilation field in the magnetic nanodots  we started with a model proposed by Guslienko {\it et al.} \cite{Guslienko1:2001,Guslienko2:2001} to investigate the vortex behavior in submicron  dots. These authors considered a ferromagnetic dot with a height $L$ and a radius $R$ that presents  a vortex state with a distribution of the unit magnetization in cylindrical coordinates $\rho,\varphi,z$  given by $\vec{m}= \sin \theta \left( \rho \right) \hat{\phi}+ \cos \theta \left( \rho \right) \hat{z}$, where\cite{Guslienko2:2001}
\begin{equation}
m_{\phi}= \sin \theta \left( \rho \right)=\left\{ \begin{array}{cc}
(2b\rho / \left(b^2+\rho^2\right))  & \rho \leq b\\
1 & \rho \geq b \end{array}\right.\, .
\label{lande11}
\end{equation}
\noindent
Here $b$ is the radius of the core. If we consider magnetostatic, exchange, and Zeeman contributions to the energy,  the normalized dimensionless vortex annihilation field in the rigid core model proposed by  Guslienko {\it et al.}\cite{Guslienko2:2001} is written as
\begin{equation}
h_{an}\left(\beta, R\right)=4\pi F_1\left(\beta \right)-\left(\frac{R_0}{R}\right)^2\,,
\label{annihi}
\end{equation}
\noindent
where $\beta=L/R$, $R_0$ is the exchange length and $F_1\left(\beta \right)$ is given by

\begin{equation}
F_{1}\left(\beta, R\right)=\int_0^\infty  \left(1-\frac{1-e^{-\beta t}}{\beta t }\right)J_1^2\left(t\right) \frac{dt}{t}\,.
\end{equation}
 \begin{figure}
 \includegraphics[width=1\columnwidth]{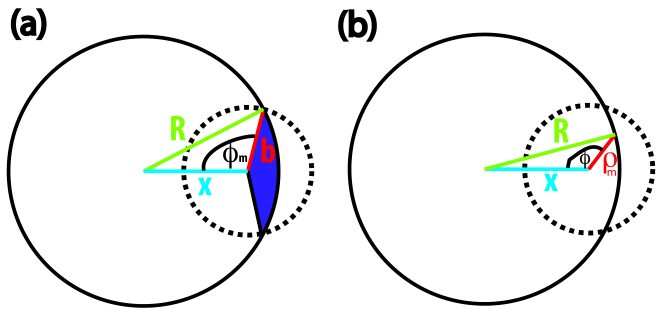}\caption{Geometrical relation between the vortex core, defined by the dotted line,  and the full  dot.  {\bf a)} Illustration of the angle $\phi_m$ that depends on the radius of the dot, $R$, radius of the core, $b$, and separation between the centers of the dot and core, $x$. {\bf b)} Representation of $\rho_m$,  that depends on $R$,  $x$, and the angle $\phi$ between $x$ and $\rho_m$.}
\label{figura1}
\end{figure}

\subsection{Introducing a Perpendicular Uniaxial Anisotropy }

While the model proposed by Guslienko {\it et al.} \cite{Guslienko1:2001,Guslienko2:2001} contains no anisotropy, in our calculations we include a uniaxial anisotropy along the ${z}$ axis and focus on its effect on the annihilation field.
We start calculating the anisotropy energy contribution of the system that is given by
\begin{equation}
W_{K}=- L K_z \int \left(\vec{m} \cdot \hat{z} \right)^2 \rho d\phi d\rho\,,
\label{wan}
\end{equation}
\noindent
where $K_z>0$ is the anisotropy constant and $ \hat{z}$ is the easy axis. From this expression, the contribution to the energy due to the anisotropy comes only from the core region inside the dot.  From Fig. \ref{figura1} we obtain
\begin{equation}
\phi_m=\arccos \left( \frac{x^2+b^2-R^2}{2 x b}\right) \, ,
\label{wan2}
\end{equation}
\noindent
and
\begin{equation}
\rho_m=x \cos \phi + \sqrt{R^2-x^2+x^2 \cos^2 \phi} \ .
\label{wan3}
\end{equation}
Using these expressions  we can write  Eq. \ref{wan} as

\begin{equation}
\begin{split}
W_{K}=-2 K_z L\int_{0}^{\phi_m}\left[ \int_0^b m_z^2\rho d\rho \right]d\phi\\
-2 K_z L\int_{\phi_m}^{\pi}\left[ \int_0^{\rho_m} m_z^2\rho d\rho \right]d\phi
\end{split}
\end{equation}

\begin{equation}
\begin{split}
W_{K} = -K_z Lb^2 \sec^{-1}\left[\frac{2 b x}{b^2 - R^2 + x^2}\right] \left(3 - 2 \ln4 \right)
-G\,.
\label{eq:}
\end{split}
\end{equation}
%\noindent
In this expression $m_z^2= \left(1-4 b^2 \rho^2/\left(b^2+\rho^2\right)^2\right)$ and $G$ represents the contributions to the anisotropy energy shown in the dark regions in Fig. \ref{figura1}a
\begin{equation}
G=2 K_z L \int_{\phi_m}^{\pi}\left[  \int_0^{\rho_m} \left(1-\frac{4 b^2 \rho^2}{\left(b^2+\rho^2\right)^2}\right) \rho \, d\rho \right]d\phi \ .
\label{eq2}
\end{equation}

When $\phi_m\left(x\rightarrow R\right)\approx \pi/2$ or $c=b/R \ll 1$,  $G$ can be approximated to zero at first order of $\left(R-x\right)$. However, in our calculations we considered it explicitly. If the anisotropy energy is normalized to $M_s^2 V$, that is, $w_{K}=W_{K}/(M_s^2 V)$, and using $s=x/R$,  $c=b/R$ and $V=\pi R^2 L$, we obtain

\begin{equation}
\begin{split}
w_{K}\left(s\right)
=\frac{-K_z c^2}{\pi M_s^2} \sec^{-1}\left[\frac{2 c s}{c^2 - 1 + s^2}\right] \left(3 - 2 \ln4 \right)\\
-g\left(s\right),
\end{split}
\label{eq3}
\end{equation}
\noindent
where $g\left(s\right)=G/M_s^2 V$. We proceed by minimizing the magnetic anisotropy energy with respect to $s$ and evaluating in the equilibrium displacement where the vortex center reaches the dot perimeter. In other words, differentiating Eq. \ref{eq3} with respect to $s$ and taking the limit $s\rightarrow 1$, we obtain the value of the contribution of the anisotropy  to the annihilation field

\begin{equation}
\begin{split}
h_{K} = \lim_{s\rightarrow 1}\frac{\partial w_{K}\left(s\right)}{\partial s} \\
=-\frac{K_z}{ M_s^2} \frac{c \left(c^2 - 2\right) \left( \ln16 - 3\right)}{\pi \sqrt{4 - c^2}}
- \lim_{s\rightarrow 1}\frac{\partial g\left(s\right)}{\partial s}\,.
\label{eq4}
\end{split}
\end{equation}

In this way, and adding this expression to the annihilation field given by Eq. \ref{annihi}, we obtain  the annihilation field for a nanodot with perpendicular anisotropy
\begin{equation}
\begin{split}
h_{an}\left(\beta, R\right)=4\pi F_1\left(\beta \right)-\left(\frac{R_0}{R}\right)^2 \\
-\frac{K_z}{ M_s^2} \frac{c \left( c^2 - 2\right) \left(-3 + \ln16\right)}{\pi \sqrt{4 - c^2}}
- \lim_{s\rightarrow 1}\frac{\partial g\left(s\right)}{\partial s}\,.
\end{split}
\end{equation}

\section{Results and Discussion}
\label{Sec:Results}
Our analyses for the annihilation field are based on a theoretical approach and on micromagnetic simulations. From both approaches we have obtained the dependence with the perpendicular anisotropy of the annihilation fields; from the simulations the core diameters were also obtained. We present below the results for the two cases: {\ref{sub:Annihilation} annihilation field, and {\ref{sub:CoreShape} evolution of the magnetic core shape. We have observed that the analytical calculations result in larger annihilation fields as compared with the numerical simulations, however, there is a qualitative agreement between the results from both models.  A qualitative agreement between theretical results and measured annihilation fields was previously reported.\cite{Novosad:2001}
\begin{figure}[h]
\centering
{\includegraphics[width=1\columnwidth]{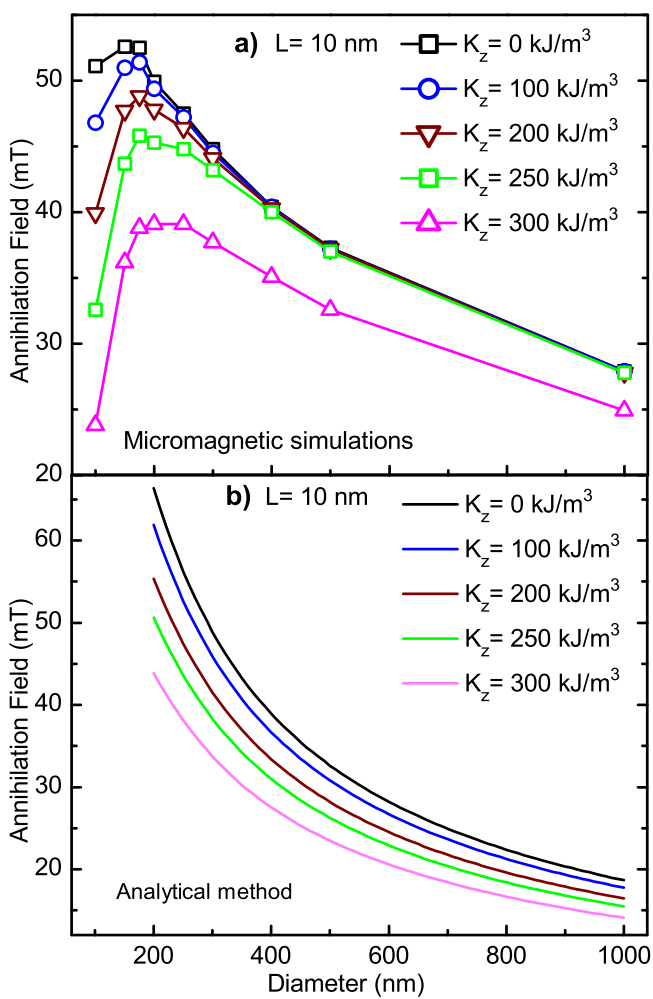}}
\caption{(Color Online) Annihilation fields versus diameter for dots {\bf a)} obtained by micromagnetic simulation, and {\bf b)} obtained by analytical method. These graphs show the influence of the value of the perpendicular anisotropy on the annihilation field for different diameters.}
\label{Fig:10AnnihilationXDiam}
\end{figure}

%===========
\subsection{Annihilation fields}
\label{sub:Annihilation}
The values of the annihilation fields $B_{an}$ as a function of disk diameter, for different anisotropies $K_z$, obtained from both methods are given in Figs. \ref{Fig:10AnnihilationXDiam}, \ref{Fig:20AnnihilationXDiam} and \ref{Fig:30AnnihilationXDiam} for $L=10, \ 20 \ {\rm and} \ 30$\,nm, respectively.

%\textcolor{red}{In the thin disk limit, one finds that the agreement between the values of $\Delta B_{an}$ obtained from the analytical method and from the simulations is good (see Fig. \ref{Fig:ComparacaoThin}, where the %values of $\Delta B_{an}$ obtained from both methods are given as a function of the diameter, for different values of $K_z$).  The values of $\Delta B_{an}$ obtained in both methods also shows the same trend, but %quantitatively there is no agreement. The differences between the two sets of results are larger for small disks diameters, as shown in Fig. \ref{Fig:comparacao}, where $K_z^{max}= 225,\, \rm{and} \, 165$ kJ/m$^3$ for %the thicknesses $L=20 \, \rm{and} \, 30$ nm.}  \textcolor{blue}{ This agreement suggests that the theoretical model gives good results in the thin disk limit. The best agreement is found for disk diameters larger than 200 %nm; this can be explained from the fact that the rigid vortex model is calculated in the limit where the vortex core diameter is much smaller than the disk diameter.}
%\textcolor{red}{ A similar effect can be observed by plotting the difference $\Delta B_{an}=B_{an}(K_z=0)-B_{an}(K_z=K_z^{max})$ where $K_z^{max}$ is the maximum value of the anisotropy that is consistent with a %magnetic vortex structure.}

%The qualitative behavior observed in the analytical results for $B_{an}$ and the micromagnetic simulations generally match, both in the overall tendency and 
The two sets of results agree in the fact that, as the diameters of the disks increase, the values become less dependent on the anisotropy (Figs.~\ref{Fig:10AnnihilationXDiam},~\ref{Fig:20AnnihilationXDiam} ~and~\ref{Fig:30AnnihilationXDiam}).

%This shows that the results of the theory and simulations agree qualitatively for all thickness.
\begin{figure}[h]
\centering
{\includegraphics[width=1\columnwidth]{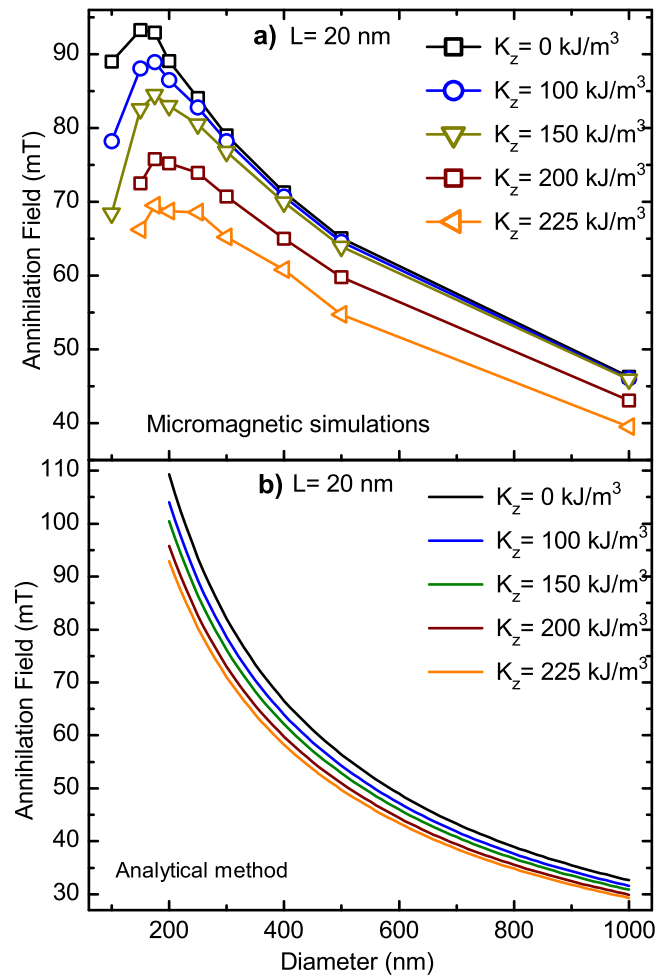}}
\caption{(Color Online) Annihilation fields versus diameter for dots {\bf a)} obtained by numerical calculation and  {\bf b)} obtained by the analytical method.}
\label{Fig:20AnnihilationXDiam}
\end{figure}

Increasing the anisotropy, which results in larger core sizes, one is led to lower annihilation fields. 
%From Figs. \ref{Fig:10AnnihilationXDiam}, \ref{Fig:20AnnihilationXDiam} and \ref{Fig:30AnnihilationXDiam}  we can see that analytical calculations result in larger annihilation fields as compared with numerical simulations. In this way, both methods show the influence of $K_z$ on $B_{an}$.

For small diameters, the effect of the anisotropy is more noticeable than for the large disks, both in the theory and simulations. In the simulations, the sensitivity of the annihilation field to changes in anisotropy increases with the value of $K_z$.

The disagreement between the two methods can be related to a deformation of the core observed in the micromagnetic simulations. 
\begin{figure}[h]
\centering
\includegraphics[width=1\columnwidth]{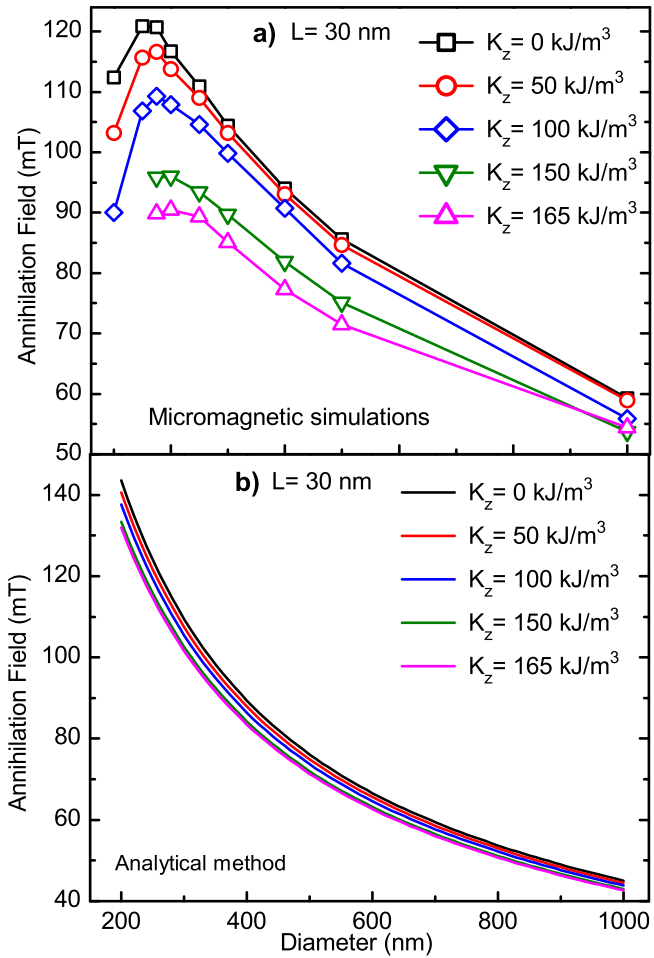}
\caption{(Color Online) Annihilation fields versus diameter for dots with {\bf a)} and obtained by numerical calculation and  {\bf b)} obtained by the analytical method.}
\label{Fig:30AnnihilationXDiam}
\end{figure}

\subsection{Evolution of the magnetic core shape}
\label{sub:CoreShape}
In the search of the effect that gives rise to the disagreement between the values of $B_{an}$ derived by analytical method and by numerical calculation, as we mentioned above, we considered the possibility that this effect would also cause a deformation of the vortex cores. Our micromagnetic simulations show that this deformation is in fact present, as shown in Figs. \ref{Fig:profile-h10d500Extreme} and \ref{Fig:profile-h30d500Extreme}. Of course, the rigid vortex model breaks down in the cases where important deformations of the vortex core are observed.

We have therefore analyzed the evolution of the shape of the magnetic vortex core as it moves towards the edge of the disks, under the influence of an applied magnetic field; the effect of varying the value of a perpendicular anisotropy $K_z$ was also investigated. The simulation results show a gradual deformation of the vortex cores, that can be quantified; the vortex cores change from a circular shape to a nearly elliptical (``banana-like") shape. Figs.~\ref{Fig:profile-h10d500Extreme}~and~\ref{Fig:profile-h30d500Extreme} show images of the vortex core and the core profile for $L=10$ nm and $L=30$ nm; the change in shape of the vortex core is very clear.

Comparing Fig.~\ref{Fig:profile-h10d500Center} and Figs.~\ref{Fig:profile-h10d500Extreme}~and~\ref{Fig:profile-h30d500Extreme}, we observe that the vortex core deformation increases as the core moves away from the center. The deformation is zero (i.e., the core is circular) for $B=0$ and maximum when the vortex core is on the edge of the disk, in the instant immediately before the vortex annihilation.

To clarify this point we have drawn the magnetization profile of the core for fields close to the annihilation field for  dots of $L=10$ and 30 nm, as shown in Figs. \ref{Fig:profile-h10d500Extreme}a and  \ref{Fig:profile-h30d500Extreme}a, respectively. The dark region (blue online) in these figures represents the core region. The profile of the vortex core at the center of the disk is shown in Fig. \ref{Fig:profile-h10d500Center}. The Figs. \ref{Fig:profile-h10d500Center}a, \ref{Fig:profile-h10d500Center}b and \ref{Fig:profile-h10d500Center}c show the effect of the anisotropy on a dot of diameter $500$ nm with $L=10$ nm, and Figs. \ref{Fig:profile-h10d500Center}d, \ref{Fig:profile-h10d500Center}e and \ref{Fig:profile-h10d500Center}f  for diameter $500$ nm, with $L=30$ nm.

For $L=10$\,nm the core keeps a nearly circular shape, the core deformation reaches about 10 \% for zero anisotropy ($K_z = 0$), and for  $K_z = 300$\,kJ/m$^3$ the deformation is around 30\%. Comparing Figs. \ref{Fig:profile-h10d500Center}a, \ref{Fig:profile-h10d500Center}b and \ref{Fig:profile-h10d500Center}c with Fig. \ref{Fig:profile-h10d500Extreme}, it is evident the vortex core deformation in the latter, due to an increase of its size along the $y$ axis and the depth of the magnetization dip.
\begin{figure}[h]
\centering
\includegraphics[width=1\columnwidth]{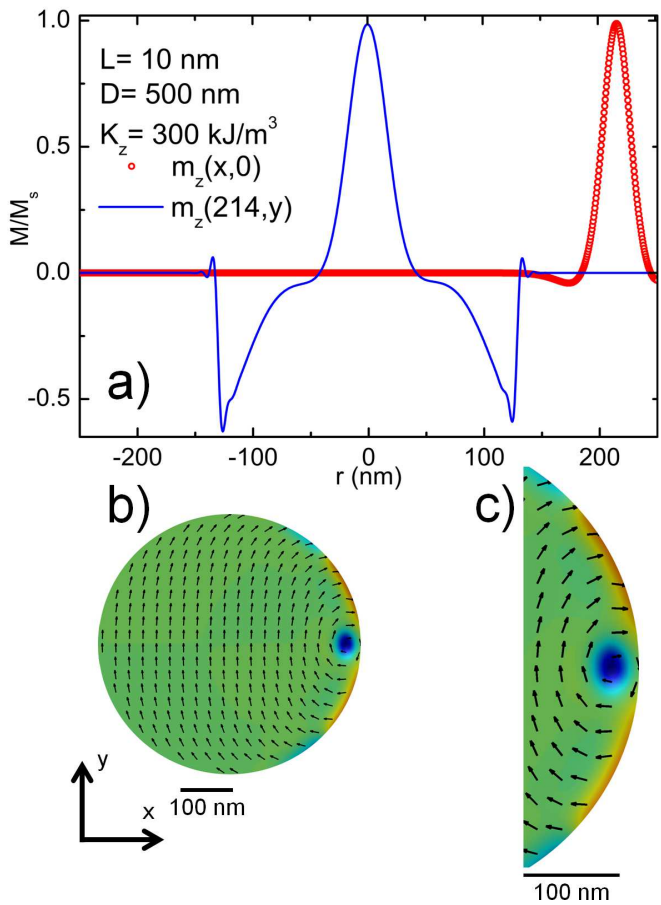}
\caption{(Color Online) Change of the shape of the vortex core for $L=10$\,nm, $D=500$\,nm and $K_z=300$\,kJ/m$^3$  immediately before the annihilation. {\bf a)} profile of the core along the $x$ axis (red dotted line) and along the $y$ axis (blue continuous line). {\bf b)} representation of the disk. {\bf c)} a detail of the disk section close to the vortex core.}
\label{Fig:profile-h10d500Extreme}
\end{figure}

However, for $L = 30$\,nm, the smallest core deformation is 30\%, that corresponds to anisotropy  $K_z = 0$, and a larger core deformation is observed for $K_z = 165$\,kJ/m$^3$, of about 100\%. Therefore, the core loses its circular shape for large anisotropy constant values $K_z$, leading to a core of roughly elliptical section that is not well described by the rigid vortex model. This is shown in Figs. \ref{Fig:profile-h10d500Extreme} and \ref{Fig:profile-h30d500Extreme}, where it is more evident the deformation of the vortex core, as well as the variation in the magnetization dip. It is important to note that a higher anisotropy constant for $L = 30$ nm will result in a magnetic configuration that is not a vortex, as observed above. Therefore $K_z = 165$\,kJ/m$^3$ is the largest value that can be considered in order to obtain a vortex state.
%In these cases the volume occupied by the core in the analytical calculations is much smaller in the %frame of the rigid vortex model, resulting in higher annihilation fields as compared to the simulations.
\begin{figure}[h]
\includegraphics[width=1\columnwidth]{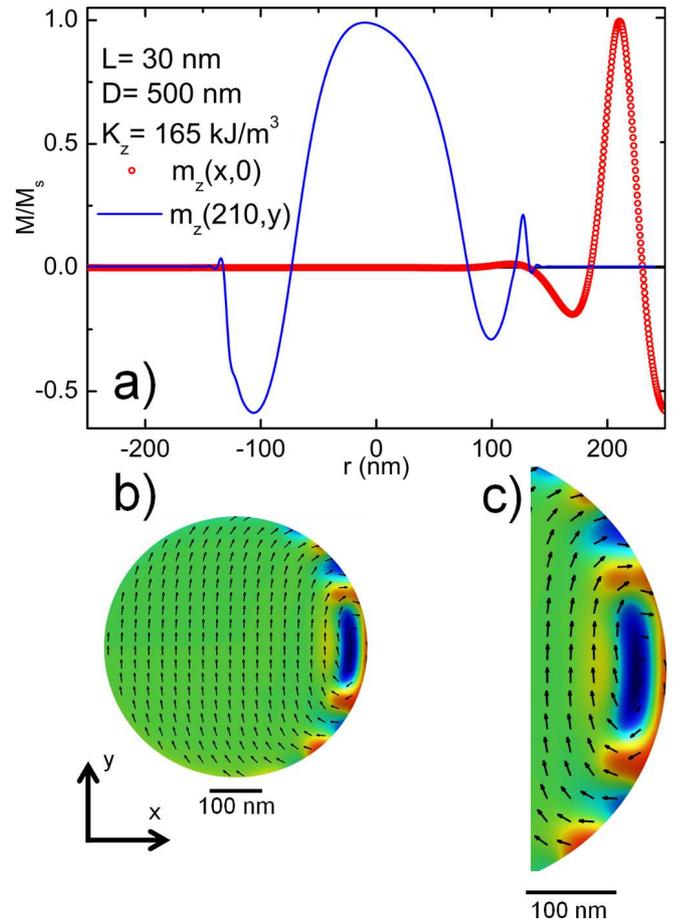}
\caption{(Color online) Change of the shape of the vortex core for $L=30$\,nm, $D=500$\,nm and $K_z=165$\,kJ/m$^3$ at positions immediately before annihilation.
In {\bf a)} profile of the core along the $x$ axis (red dotted line) and along the $y$ axis (blue continuous line), in {\bf b)} image of the disk and c) a detail of the disk.}
\label{Fig:profile-h30d500Extreme}
\end{figure}

We observe that the core deformation is not present for ${\mathbf B =0}$, as shown in Fig. \ref{Fig:profile-h10d500Center}. However, as the field begins to increase, the process of deformation of the core sets in, as can be seen in Figs. \ref{fig:k000deformationXCore} and \ref{fig:k100deformationXCore}, plotted from values obtained using Eq. (\ref{eq:deformation}); they are shown as a function of relative core positions ($p_{core}/R$), defined as the ratio of the distance of the core center from the center of the disk ($p_{core}$) divided by the disk radius ($R$).
\begin{figure}[h]
\includegraphics[width=1\columnwidth]{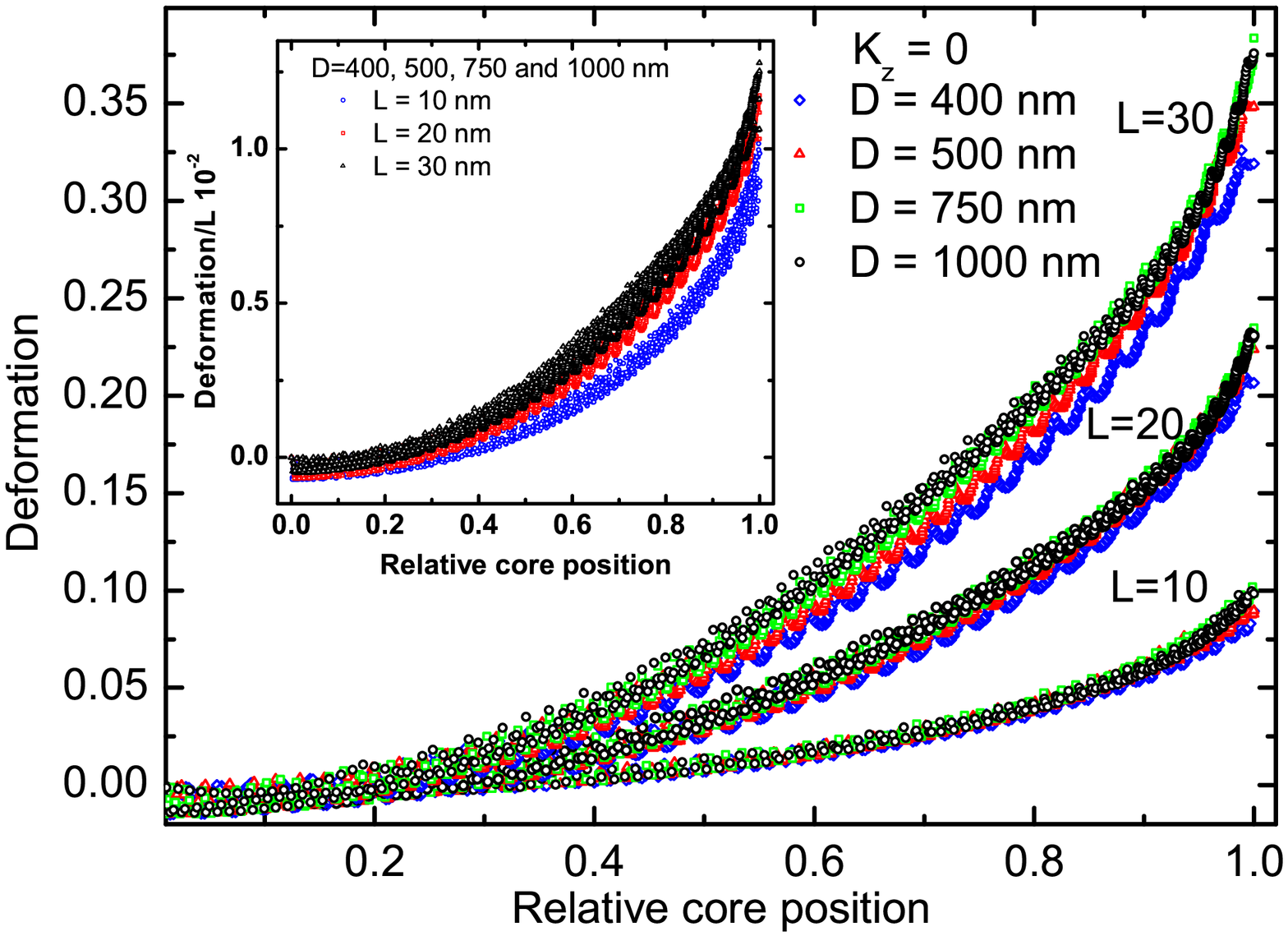}
\caption{(Color online) Deformation $\delta =(r_y-r_x)/r_x$ for $K_z=0$, $L=10, \, 20, \, \rm{and}\, 30$\,nm for diameters $D=~400, \, 500, \, 750 \, \rm{and} \, 1000$\,nm versus normalized core position ($p_{core}/R$). Note that a scaling law is apparent.}
\label{fig:k000deformationXCore}
\end{figure}

Figures \ref{fig:k000deformationXCore} and \ref{fig:k100deformationXCore} show different stages of deformation. In the case where the vortex core is located below 25\% of the radius of the disk, the  deformation can be neglected. From this point onwards the deformation begins to increase, and the maximum is reached when the core approaches the edge of the disk. Note in Figs. \ref{fig:k000deformationXCore} and \ref{fig:k100deformationXCore} that for each thickness (10, 20 and 30 nm) four curves were plotted for different diameters (400, 500, 750 and 1000 nm) and that these curves overlap; it appears to exist a scaling law for the deformation of the core, i.e., for each disk thickness, the deformation does not depend on the disk diameter, it depends only on the relative position of the vortex core.

\begin{figure}[h]
\includegraphics[width=1\columnwidth]{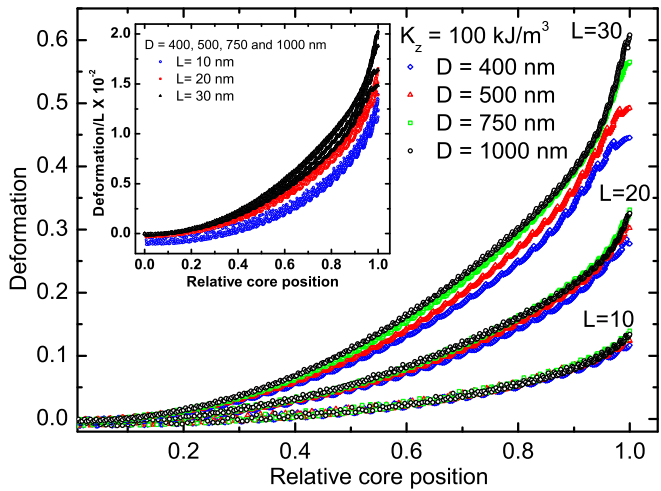}
\caption{(Color online) Deformation $\delta =(r_y-r_x)/r_x$ for $K_z=100$\,kJ/m$^3$, $L=10, \, 20, \, \rm{and} \, 30$\,nm for diameters $D~=~400, \, 500, \, 750 \, \rm{and} \, 1000$\,nm versus normalized core position ($p_{core}/R$). Note that a scaling law is apparent.}
\label{fig:k100deformationXCore}
\end{figure}

\section{Conclusions}
\label{Sec:Conclusions}
In summary, by means of an analytical model and numerical simulations we have obtained the annihiliation fields for dots of different heights and anisotropy constants. In all cases, the annihilation fields decrease with increasing anisotropy constant $K_z$ and with increasing disk diameter. The values of $K_z$ and disk height $L$ have a stronger effect on the annihilation fields of the smaller disks. However, the anisotropy and the thickness operate in an inverse way on the annihilation field; whereas the increase in height increases the annihilation field, from the analytical results using the rigid vortex model, the increase in anisotropy decreases this field. Finally, we have shown the variation in the deformation of the vortex core $\delta$ as a function of the perpendicular anisotropy and disk height. The occurrence of this deformation evidenced in the micromagnetic simulations suggests that it has to be taken into account in the description of the dynamics of the magnetic vortices. The deformation of the core does not scale with the radius of the disks, it is only related to the relative position of the core.

\section*{acknowledgments}
In Brazil we acknowledge the support of the agencies CAPES, CNPq, FAPERJ and FAPESP. In Chile we acknowledge the partial support from FONDECYT under grants  11121214, 1120356 and 1120618, from the
Center for the Development of Nanoscience and Nanotechnology, and from  ICM P10-06-F funded by Fondo de Innovaci\'on para la Competitividad, from the MINECON.

%\bibliography{ArtAnniNucle}


\end{document}